\documentclass[a4paper,twoside]{article}

\usepackage{epsfig}
\usepackage{subcaption}
\usepackage{calc}
\usepackage{amssymb}
\usepackage{amstext}
\usepackage{amsmath}
\usepackage{amsthm}
\usepackage{multicol}
\usepackage{pslatex}
\usepackage{apalike}
\usepackage{algorithm2e}
\usepackage{url}
\usepackage[bottom]{footmisc}
\usepackage{hyperref}
\usepackage{listings}
\usepackage{protobuf/lang}
\usepackage{protobuf/style}
\usepackage{todonotes}
\usepackage{courier}
\usepackage{SCITEPRESS}     
\usetikzlibrary{tikzmark}

\newcommand{\overlaycircle}[2]{%
	\tikz[overlay,baseline=(char.base)]\node[anchor=north, draw=#1, circle, inner
	sep=1pt,fill=#1,text=white, scale=0.9, minimum size=13pt, inner sep=0pt,font=\sffamily,
	solid](char){#2} ;}
\newcommand{\circled}[2]{%
	\tikz[baseline=(char.base)]\node[anchor=north, draw=#1,circle, inner
	sep=1pt,fill=#1,text=white, scale=0.9, minimum size=13pt, inner sep=0pt,
	solid](char){#2} ;}

\begin{document}
\title{owl2proto: Enabling Semantic Processing in Modern Cloud Micro-Services}

\author{\authorname{Christian Banse\orcidAuthor{0000-0002-4874-0273}, Angelika Schneider\orcidAuthor{0000-0002-8962-3276} and Immanuel Kunz\orcidAuthor{0000-0002-4669-0030}}
\affiliation{Fraunhofer AISEC, Garching b. Muenchen}
\email{\{firstname.lastname\}@aisec.fraunhofer.de}
}


\abstract{The usefulness of semantic technologies in the context of security has
been demonstrated many times, e.g., for processing certification evidence, log
files, and creating security policies. Integrating semantic technologies, like
ontologies, in an automated workflow, however, is cumbersome since they
introduce disruptions between the different technologies and data formats that
are used. This is especially true for modern cloud-native applications, which
rely heavily on technologies such as protobuf. In this paper we argue that these
technology disruptions represent a major hindrance to the adoption of semantic
technologies into the cloud and more effort and research is required to overcome
them. We created one such approach called \textit{owl2proto}, which provides an
automatic translation of OWL ontologies into the protobuf data format. We
showcase the seamless integration of an ontology and transmission of semantic
data in an already existing cloud micro-service.}

\onecolumn \maketitle \normalsize \setcounter{footnote}{0} \vfill

\section{\uppercase{Introduction}}
\label{sec:introduction}

Semantic technologies can establish a common understanding of, e.g., cloud
concepts and their properties and thus have a high importance for the
interoperability of cloud services. In the security context, semantic
technologies have, for instance, been used to model certification evidence
\cite{banse2021cloudpg,banse2023semantic}, to structure information in log files
\cite{benshimol2024observability}, or to model general cloud security concepts
\cite{takahashi2010ontological}. 

We argue that while the academic discussion focuses on the semantic design of,
e.g., cloud security concepts \cite{maroc2019comparative}, the technological
integration of semantic design and its technological implementation is lagging
behind. Modern technologies like micro-services and RPCs, for example, are not
integrated with technologies of the semantic web stack. We think that a better
integration with such technologies would make semantic concepts easier to use,
increase its adoption in different domains, and it could improve the
interoperability of cloud systems, e.g. in multi-cloud and cloud-edge scenarios.



In this paper, we focus on \textit{protobuf} as an example for this position and
use the cloud security context as an example application domain. Note, however,
that our arguments apply beyond these examples. Protobuf is one of the most
commonly used technologies for micro-services. Originally designed as a format
to describe the serialization of network packages, it has evolved into an
interface definition language, not only describing the exchanged data, but also
the services that produce or consume this data. 

Protobuf intentionally does not focus on the semantics of the exchanged data.
Instead, it defines a syntax and structure of an object (called message), by
describing which fields a programmer would use to fill this object. This
includes primitive types, arrays and other messages. But the semantics of the
data, such as, that it describes evidences gathered for a security incident, is
beyond its scope. Developers would have to resort to storing this semantic
information in other formats (e.g. RDF, JSON-LD, etc.) and then transmitting the
actual data in a serialized form, creating a technology gap between the
``semantic'' world and the rest of the application.

In this paper, we argue for bridging the gap between semantic technologies and
the integration with modern data processing. We demonstrate how to advance this
integration by introducing a methodology and implementation that transforms
ontology concepts into RPC definitions, and we point out use cases. The
implemented tool is called \textit{owl2proto}.


\section{Background and Related Work}

This section contains an overview of related work on semantic data processing,
especially in the context of security and privacy. Additionally, we provide a
background on the protobuf format our approach is based on.

\subsection{Processing of Semantic Data}
\label{sec:previous}

The term ``semantic''---in the context of IT technology---refers to labeling
data with common tags to enable their automatic recognition and processing.
Ontologies provide one way to encode relationships between concepts, for example
to represent hierarchies between them or other types of relationships. They can
then be used by automatic technologies to put words, source code, and other
objects into the right context. Several technologies exist to serialize such
semantic data, e.g. to send them across communication channels. These include
simple RDF serialization \cite{rdf11}, XML serialization of OWL \cite{owl2} or
more complex protocols such as JSON for Linking Data (JSON-LD)
\cite{sporny2020json}.

In the context of security and privacy, semantic technologies have been used for
malware analysis~\cite{carvalho2016malware}, authentication and
authorization~\cite{servos2017current}, and in governance
engineering~\cite{esteves2024analysis,nadal2022operationalizing}. Even systems
to automate the adherence to security compliance frameworks have been
proposed~\cite{banse2021cloudpg,banse2023semantic}.

\subsection{gRPC and protobuf}
\label{sub:protobuf}


Protocol Buffers (or protobuf in short) is an open-source project to handle and
transmit binary data in a structured way. It was  originally developed
internally in Google and released to the public in 2008. Protobuf's purpose is
two-fold. First, it defines a binary wire format for the transmission of
arbitrary data (so called \textit{Messages}). It is very efficient and therefore
used in scenarios where a high throughput of data is required. Second, protobuf
can be seen as a \textit{Interface Definition Language (IDL)}, which describes
data structures and RPC interfaces in a (programming)-language independent way. 

These IDL files can then be used to auto-generate appropriate data structures in
many chosen programming languages such as C/C++, Java, Python, Go and others.
Additionally, code for exposing the RPC interfaces as a service and consuming
these as a client can also be auto-generated. Therefore, it is often used in the
communication between a mesh of micro-services, also in combination with
additional RPC frameworks such as gRPC\footnote{\url{https://grpc.io}} or
Connect\footnote{\url{https://connectrpc.com}}, which take care of the actual
network transmission of protobuf messages.

\begin{lstlisting}[escapechar=!,caption={A protobuf message displaying some of the core concepts, 
  such as messages, message options as well as the \textit{oneof} keyword.},
  language=protobuf3, style=protobuf, label=lst:proto]
message MyMessage {  !\overlaycircle{black}{1}!
!\overlaycircle{gray}{1a}!  option (note) = "important";

  int32 id = 1;
  optional string name = 2;
!\overlaycircle{black}{2}!  MyOtherMessage other = 3;
  
!\overlaycircle{black}{3}!  oneof alternatives {
!\overlaycircle{gray}{3a}!    string good_alternative = 10;
!\overlaycircle{gray}{3b}!    bool bad_alternative = 11;
  }
}

message MyOtherMessage {
  string another_name = 1;

  reserved 2;  !\overlaycircle{black}{4}!
}

extend google.protobuf.MessageOptions {  !\overlaycircle{black}{5}!
  repeated string note = 100;
}
\end{lstlisting}

\autoref{lst:proto} demonstrates several core concepts of the protobuf language
as. First, every data structure in protobuf is called a \textit{Message}
(\circled{black}{1}). Each message must have a unique name (per package) and
contains a list of fields. Each field also has a name, a type and a field number
(the number after the equals sign). The type can either be one of the defined
scalar types (such as \texttt{string}, \texttt{int32} and
others\footnote{\url{https://protobuf.dev/programming-guides/proto3/}})
or another message (\circled{black}{2}). The name as well as the field number
must be unique inside their respective message. When protobuf messages are
serialized and de-serialized, only field numbers are transmitted. Therefore, it
is of utmost importance that field numbers are never changed or re-used for
other fields, as long as compatibility needs to be ensured (e.g., within one
major version). 

To support this, the keyword \texttt{reserved} (\circled{black}{4}) can be used
to denote that there was once a field with a certain number (2 in this case)
which is not used and cannot be used anymore.
Protobuf also supports a special construct called \texttt{oneof}. It is somewhat
similar to a \textit{union} data structure in C. It defines a subgroup of fields
(\circled{black}{3}), in which only one of the defined fields
(\circled{gray}{3a} or \circled{gray}{3b}) can be set at a given time. 

Finally, the notion of options can be seen as a system similar to annotations in
other languages. They can provide additional meta-data to files, messages or
fields. The contents of these are not transmitted over the wire, but they are
available to both the transmitting and receiving party in the generated code.
Next to a list of built-in options, also new options can be declared, as seen in
\circled{black}{5}. In this example, a new option \texttt{note} is declared
which can be used to annotate a message (\circled{gray}{1a}).


\section{Motivation and Use Case}


\subsection{Use Case: Collecting Semantic Information for Cloud Security}
\label{sec:usecase}

\begin{figure}  
  \includegraphics[width=\columnwidth,trim={0.5cm 0 0.5cm 0},clip]{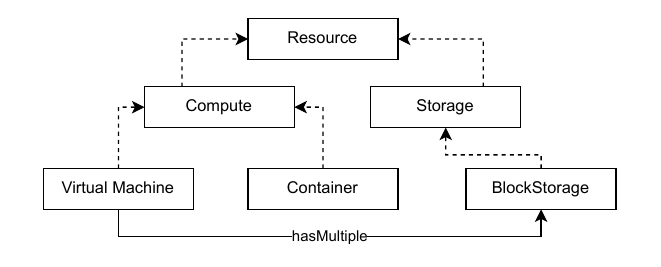}
  \caption{Small excerpt of an ontology of cloud resources based on
  \cite{banse2023semantic}}
  \label{fig:ontology}
\end{figure}

To motivate the need for our approach, we considered existing previous works
(see \autoref{sec:previous}) from the security field that process semantic in
the cloud. In the following, we will consider the design of
\cite{banse2023semantic} because they are already leveraging cloud technologies
such as gRPC in addition to semantic data, but did not fully bridge the gap
between both. They propose the collection of semantic information (also called
\textit{evidence}) about a cloud system, such as virtual resources and their
properties. The aim is to assess certain properties of these assets with respect
to security or privacy -- in the context of certification. Because of the
potential large amount of data gathered, their proposed system is split into
different micro-services:
\begin{itemize}
  \item a group of services (\textit{discovery}) is responsible for collecting
  information about a cloud service and putting them into data objects described by
  an ontology (see \autoref{fig:ontology} for a small excerpt),
  \item a group of services (\textit{evidence store}) is storing the gathered
  information in a common database,
  \item a group of services (\textit{assessment}) is responsible for comparing
  the properties of such an evidence against a specific set of desired states
  using a rule engine.
\end{itemize}

All services use modern communication protocols, such as gRPC to communicate
with each other, because of the high throughput needed. protobuf is also
used to model and describe the service itself.

\begin{lstlisting}[language=protobuf3, style=protobuf,label=lst:assessment,caption=Excerpt of the assessment service API described in \cite{banse2023semantic}]
message Evidence {
  string id = 1;
  google.protobuf.Struct evidence = 2;
}

message AssessEvidenceRequest {
  Evidence evidence = 1;
}

service Assessment {
  rpc AssessEvidence(AssessEvidenceRequest) 
  returns (AssessEvidenceResponse);
}
\end{lstlisting}

\autoref{lst:assessment} shows a very small excerpt of this API definition,
detailing the \textit{Assessment} service, whose primary RPC call takes in an
\textit{Evidence}. While the evidence itself is described within the discovery
services in terms of an ontology, the listing shows that this semantic
information is partially lost. It is simply translated as a \textit{Struct},
which is a protobuf definition of a generic key-value store. While this keeps
the basic information -- similar to a simple JSON -- all other semantic aspects,
such as entity inheritance or links to specific ontologies are lost. Instead a
better approach would be dedicated messages for ontology types, like
\textit{Compute}, \textit{VirtualMachine} and so on. Overall, it makes it very
hard for users of this API because they need to have a look at the protobuf
definition and additionally analyze how to model the contents of
evidence/resource.

The only way to transmit such data and not loose semantic information would be
to use formats like JSON-LD \cite{sporny2020json} or other variants of JSON that
are enriched with semantic information. This has several shortcomings (which are
discussed in more detail in \autoref{sec:jsonld}). Therefore, this service
would benefit from a homogeneous solution in which the semantic data would also
be available in the protobuf schema.

\subsection{Generalization of the Use Case}

Numerous similar cases can be found where differently named---but semantically
similar---data need to be abstracted to the same level to enable, e.g., their
aggregation. Consider the example of a social media platform for sports
activities that allows users with different types of wearables to upload their
activities including GPS coordinates, photos, their running statistics, etc.
Aggregating data from different types of hardware and software manufacturers
(e.g., smart watches, phones, specialized bike computers, etc.) implies the need
for an ontological knowledge base that abstracts the most important
properties---and tools that support the work with this knowledge base in code.
Other cases include, for instance, industrial data sharing platforms or log
ingestion and analytics.

In general, the following requirements for designing a semantic data sharing
approach for the cloud need to be considered:
\begin{itemize}
  \item \textbf{RQ1: Keeping semantic information}, such as taxonomies and
  entity inheritance during the communication between different parts of the
  system (e.g. micro-services).
  \item \textbf{RQ2: Seamless integration} with cloud native technologies, such
  as gRPC/protobuf.
  \item \textbf{RQ3: Adaptability} to the ever-changing world of DevOps and
  continuous integration and deployment.
\end{itemize}


\section{Our Approach: owl2proto}

In trying to bridge the world of cloud native frameworks and ontology design, we
choose protobuf as a means to convert ontology structures. Since protobuf is
already a schema (and service) definition, it is very well suited to describe
the exchanged data in a semantic way. We propose to auto-generate an appropriate
protobuf schema out of a modelled ontology, specifically in the Web Ontology
Language (OWL2) format. Therefore, we name our approach \textit{owl2proto}. 

\begin{figure}[h!]
  \centering
  \includegraphics[width=\columnwidth,trim={0.5cm 0 0.5cm 0},clip]{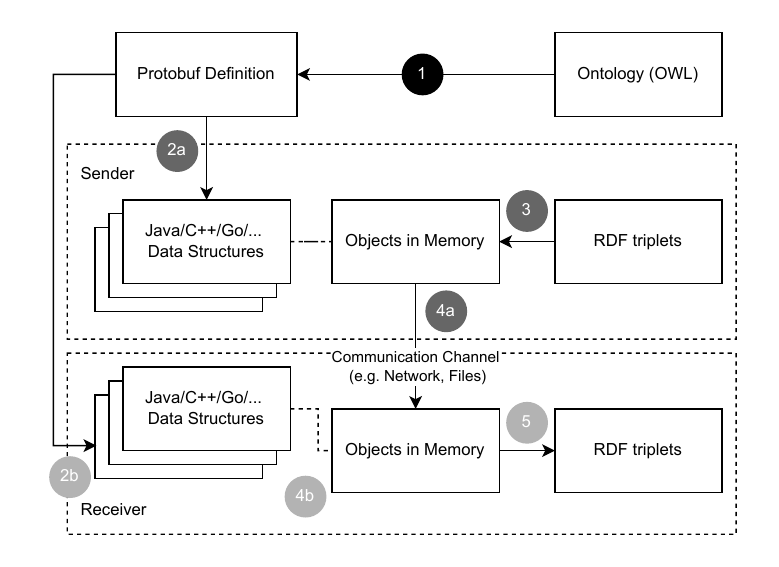}
  \caption{Our approach on how to exchange semantic data between a sender and a receiver via a channel}
  \label{fig:approach}
\end{figure}

\paragraph{Preparation phase}  \autoref{fig:approach} shows the approach in
detail. Two parties (\textit{Sender} and \textit{Receiver}) want to exchange
semantically enriched data. The first step is to auto-generate a common protobuf
definition out of the ontology that is shared by both parties
(\circled{black}{1}). Execution of the translation process (see
\autoref{sec:translate}) is only necessary once and when the ontology changes
(see \autoref{sec:sync}).

\paragraph{Semantic data exchange} Afterwards, sender (\circled{gray}{2a}) and
receiver (\circled{lightgray}{2b}) can use the regular protobuf workflow to
generate data structures in a programming language of their choice, e.g.
Java/C++/Go or any other language supported by protobuf. Since protobuf allows
the modular inclusion of different packages, it allows us to split the
auto-generated parts and the manually modelled parts of the services into
different source files. But in the end all protobuf files will be used in the
code-generation of the respective server/client and exchange objects in the
chosen programming language. The sender instantiates these data structures as
memory objects and populates them with the RDF triplets (\circled{gray}{3}) they
want to send. After the sender initiates the communication (\circled{gray}{4a}),
the receiver can de-serialize the transmitted values back into memory objects
(\circled{lightgray}{4b}). Finally, the receiver can access the transmitted RDF
triples (\circled{lightgray}{5}).


\label{sub:challenges}



\subsection{Translating Ontology Entities to protobuf Constructs}
\label{sec:translate}

One of the first steps is to map different concepts of OWL to protobuf. This
includes mapping \textbf{classes} to \textbf{messages} and \textbf{data
properties} as well as \textbf{object properties} to \textbf{fields}. We chose
the OWL format (instead of RDF) because it already includes a structure
containing classes and other concepts that are already more aligned with
concepts in other programming languages. This eases the translation to protobuf,
whereas a mapping of RDF entities would entail an additional layer of processing
--- which is essentially already done by OWL.
\clearpage 

\begin{lstlisting}[escapechar=!,language=protobuf3, style=protobuf,label=lst:example1,caption={Excerpt of a generated protobuf output based on the ontology of \autoref{fig:ontology}. A full version of the example is available in the GitHub repository.}]
option (owl.meta) = {  !\overlaycircle{gray}{0}!
  prefixes: [{
    prefix: "ex"
    iri: "http://example.com/classes"
  }, {
    prefix: "owl"
    iri: "http://www.w3.org/2002/07/owl#"
  }]};

message VirtualMachine {  !\overlaycircle{black}{1}!
!\overlaycircle{gray}{1a}!  option (owl.class).iri = "ex:VirtualMachine";
!\overlaycircle{gray}{1b}!  option (owl.class).parent = "ex:Compute";
  option (owl.class).parent = "ex:Resource";
  option (owl.class).parent = "owl:Thing";

!\overlaycircle{black}{2}!  string name = 1 [
!\overlaycircle{gray}{2a}!   (owl.property).iri = "ex:name",
!\overlaycircle{gray}{2b}!   (owl.property).parent = "owl:topDataProperty"
!\overlaycircle{gray}{2c}!   (owl.property).class_iri = "ex:Resource"
  ];

!\overlaycircle{black}{3}!  GeoLocation geo_location = 2 [
   (owl.property).iri = "ex:has",
   (owl.property).parent = "owl:topObjectProperty"
!\overlaycircle{gray}{3c}!   (owl.property).class_iri = "ex:Compute"
  ];

!\overlaycircle{black}{4}!  repeated string block_storage_ids = 3 [
   (owl.property).iri = "ex:hasMultiple",
   (owl.property).parent = "ex:has"
!\overlaycircle{gray}{4c}!   (owl.property).class_iri = "ex:VirtualMachine"
  ];
}
\end{lstlisting}

  \autoref{lst:example1} contains an excerpt of a generated protobuf output
  based on the ontology of \autoref{sec:usecase}. This illustrates the
  translation process as follows. For each OWL class, an associated protobuf
  message is created (\circled{black}{1} for the \texttt{VirtualMachine} class). 

\paragraph{Data properties}
For each data property used in the class, the (short) name of the property is
taken as the name of the field. Since each property in OWL also has a type, we
also need to map that accordingly. For primitive types, such as
\texttt{xsd:string}, we perform a simple mapping to protobuf primitive types,
such as \texttt{string} (\circled{black}{2}).


\paragraph{Object properties referring to blank nodes}
  
Properties which are only used in combination with blank (or anonymous) nodes, are
  set to the appropriate protobuf message type representing the OWL target
  class. This can be seen in \circled{black}{3}. While the OWL object property
  is named \texttt{has}, we want to name the field according to the used OWL
  class \texttt{GeoLocation} (converted to \texttt{geo\_location}). Since
  \texttt{GeoLocation} is only used in blank nodes, we can directly use the
  translated protobuf message for the field's type.

\paragraph{Object properties referring to identifiable nodes}
  Properties that refer to other nodes which definitely have their own IRI and
  are identifiable are set to the \texttt{string} datatype and an \texttt{\_id}
  suffix is added to the protobuf field name. In this case, we only store the
  IRI referring to the other node instead of the content of the node itself in
  the final protobuf message. This can be seen in \circled{black}{4}, where the
  \texttt{hasMultiple} object properties is converted to a string field of
  \texttt{block\_storage\_ids} since the target type in this class is
  \texttt{BlockStorage}\footnote{Our Prototype implementation has an internal
  mapping which object properties are translated to a singular \texttt{\_id} and
  which ones are translated to the plural \texttt{\_ids}. In the future, this
  could be taken from an annotation within the ontology.}.

\paragraph{Retaining semantic structure information}
  Protobuf message are well-suited to describe concepts like entities or OWL
  classes and their respective association to data / object properties. But, in
  order to retain the actual semantic information, such as IRIs, field types and
  other information contained in languages such as RDF or OWL, an additional
  step is needed. This entails the use of protobuf options in various forms.
  \autoref{lst:options} in the appendix contains the protobuf definition of
  these options. General information about the ontology are specified in file
  options, directly at the top of the protobuf file (\circled{gray}{0}).
  Ontology metadata about classes and properties are modelled as message
  annotations (\circled{gray}{1a}, \circled{gray}{1b}) and field annotations
  (\circled{gray}{2a}, \circled{gray}{2b}, \circled{gray}{2c}), respectively.
  Protobuf options are designed in a very efficient way and are not transmitted
  over the serialized channel. Instead the sending and receiving end can extract
  them out of fields and messages, as long as they keep their protobuf
  definitions in sync.

\subsection{Modelling Inheritance}
  Ontologies are inherently connected and use concepts like (multi)-inheritance.
  This means that a more concrete ontology entity will inherit the properties from
  all its parents. This concept does not exist in protobuf. Furthermore, some
  ontology entities (mainly non-leaf nodes) can be considered as interfaces in a
  programming language, since they hold denominators such as fields common to
  all their leaf nodes.

  Therefore, we need to flatten the hierarchy of data properties in the
    individual ontology entities when translating them to protobuf messages. For
    example, if we look at our example ontology entity \texttt{VirtualMachine},
    we can see that it derives from \texttt{Compute} and \texttt{Resource}. In
    this  case, we need to include all data properties of both parent objects as
    protobuf fields, as shown in \autoref{lst:example1} (\circled{black}{2},
    \circled{black}{3}, \circled{black}{4}). In order to keep the information
    which ancestor class specified the actual property, the option
    \texttt{(owl.property).class\_iri} is used (\circled{gray}{2c},
    \circled{gray}{3c}, \circled{gray}{4c}).

%
%
%

  Furthermore, we want to keep the class hierarchy information. We make use of
  the fact that we can specify certain message options multiple times and include
  a \texttt{(owl.class).parent} option for each ancestor (\circled{gray}{1b}). For
  non-leaf nodes we make use of the \texttt{oneof} keyword to introduce a
  message that can be used similar to an interface. 

  \subsection{Reacting to Changes}
  \label{sec:sync}
  Every time the ontology changes, a new protobuf file needs to be generated
  using \textit{owl2proto}. This step can (and should) be automated by a CI/CD
  system, such as GitHub actions. For example, a raw OWX file containing the OWL
  ontology could be stored alongside a code repository. Changes to this file can
  trigger a workflow that generates the protobuf files, so that developers can
  directly use it.

  Since protobuf is a binary protocol, it uses field numbers to differentiate
  between different data fields within a message and not names. Therefore, to be
  compatible with previous versions of a protobuf file, field numbers must not
  change or be re-used. Otherwise, transmitted data will turn up in the wrong
  field. This has two implications:
  \begin{itemize}
    \item Multiple invocations of the translation process (on the same source
    file) need to produce the same output in the same order. Especially, the
    deduction of field numbers must be deterministic.
    \item If the source file changes, all fields that exist in both the original
    as well as the modified version of the file need to also have identical
    numbers.
  \end{itemize} 

  There are several solutions to this problem, which we will discuss further,
  since each of them has their own specific drawback.
  
  \paragraph{Non-cryptographic hash}
  
  One possible solution is to derive a unique field number for each field by its
    name, e.g. through the use of a cryptographic hash, such as
    xxhash\footnote{\url{https://xxhash.com}}. But there are some caveats to
    consider:
    \begin{itemize}
      \item First, field numbers in protobuf are ranging from 1 to 536,870,911
      ($2^{29}$). The lowest xxhash implementation uses 32-bit, so we would need to
      further restrict the possible result space of hash 8 times, leading to
      more collisions.
      \item Certain field numbers (namely 19,000-19,999) are reserved for internal use
      \item Smaller numbers are more efficiently stored than larger numbers,
      contrasting a usual hash algorithm's way of using the whole result space to
      maximize entropy.
    \end{itemize}

      \paragraph{Read-in previous input}
      Another possibility is to read in the previous generated output as an
    input in order to assign existing fields the same number. For the initial
    generation, one could use a pre-defined ordering of fields based on lexical
    sorting of the field names and the parent's name which they are part of.
    Intentional space between groups of fields need to be left for
    extendability. If fields are removed, one can leverage the protobuf's
    reserved list in order to keep track of used field numbers. The greatest
    drawback of this solution is the need to have the previous version
    available, making it hard to maintain.

\subsection{Implementation}

As part of our research, we provide a prototype implementation of our approach
as an Open-Source
library\footnote{\url{https://github.com/oxisto/owl2proto}}. The
prototype is written in Go and largely uses the Go standard library to read in
OWL in its XML variant (OWX) and generates the appropriate protobuf output.

With regards to the consistently problem (see \autoref{sec:sync}), we decided to
use a non-cryptographic hash function, such as xxhash in our prototype
implementation, as this approach does not require a pre-existing proto file,
which would have required some versioning functionality. To generate the unique
field number, we use the name of the ontology class, the ontology property name
and the names of the parent classes as input for the hash function. The number
of parent classes can be arbitrary. This ensures that the field numbers remain
unique within the proto message, even if fields in different proto messages
share the same name. An example is as follows. The input values for the field
\textit{name} (cf. \autoref{lst:example1} \circled{black}{1} and
\circled{black}{2}) are a tuple of $(VirtualMachine, Compute, Resource, name)$.

Since the numerical range from 19000 to 19999 is reserved, it is imperative to
map the resulting hash value to the range of 1 to 18999. This is accomplished by
applying the mathematical modulo function. Although the numerical range from
20,000 to 536,870,911 is available for use, the range of 1 to 18,999 is adequate
for the requirements of our ontology. Additionally, smaller numbers take up less
space in protobuf serialization it is also more efficient.

\section{Discussion}

In this section we discuss whether our approach can address the requirements
that we elicited out of the use case presented in \autoref{sec:usecase} and
compare our approach to similar techniques, mainly JSON-LD.

\subsection{Addressing the Use Case Requirements}

\paragraph{RQ1: Keeping Semantic Information}
We present an approach that can translate an ontology structure into a binary
serialization format (protobuf), while keeping all the semantic information,
even when transmitting the data. This is achieved through the combination of
flattening hierarchies, translating certain types into identifiers and providing
pseudo-interfaces with \texttt{oneof} for certain entities. Finally, the use of
message and field options allow us to embed all RDF/OWL concepts like IRIs and
other annotations in the protobuf format\footnote{Although not tested, this
could potentially even allow someone to re-construct a complete OWL file out of
the generated protobuf file.}, making this requirement \textit{fully addressed}.

\paragraph{RQ2: Seamless Integration}
By choosing protobuf as our translation target, we aim at seamless integration
into the cloud native world. We can easily integrate our generated messages into
cloud native frameworks, such as gRPC or ConnectRPC since they use protobuf as
their base. In contrast to other approaches like using gRPC for the general
communication and JSON-LD for the serialization of semantic data, we can offer
one coherent API described in protobuf. This enables a more seamless developer
experience and we consider this requirement \textit{fully addressed}.

\paragraph{RQ3: Adaptability}
Arguably this is the hardest requirement to fulfill. The pace of development in
the modern cloud world can sometimes be astonishing. This also extends to the
frequency of API and data model changes. Therefore, it also stretches to the
development and enhancement of ontologies used in such services. On one side,
our approach can easily be added into a CI/CD workflow, generating new protobuf
files when ontology files change. On the other side, some inherit quirks of the
protobuf format make auto-generation of protobuf files themselves quite hard.
One such aspect is that protobuf message field numbers must not be changed once
used. This makes it very hard for approaches that auto-generate those field
numbers to a) consistently generate the same field numbers given the same input
and b) to not re-assign already used field numbers if the input changes. While
we present an initial approach using non-cryptographic hashes, our solution has
several shortcomings that need further exploration. We therefore consider this
requirement \textit{partially addressed}.

\subsection{Comparison with JSON-LD}
\label{sec:jsonld}

When using systems that already rely on JSON as a data format, JSON-LD can
provide a useful extension to increase interoperability. Regarding the technical
usability, JSON-LD has some drawbacks. For example, JSON-LD files require
dedicated libraries for processing. Furthermore, JSON-LD embeds or references
the semantic data in each transmitted message which can significantly increase
overhead. In comparison, protobuf is a binary format, which uses pre-defined
structures making it more compact and faster to serialize/deserialize. Protobuf
is therefore often considered to enable better performance.

Overall, JSON-LD can be considered an alternative to the protobuf-based approach
presented here, but it presents significant drawbacks in comparison to protobuf.
Protobuf, is the preferred choice for applications prioritizing performance and
data integrity. Its binary format and pre-defined schema ensure efficient
transmission and robust data consistency. Thus, it underlines the need for
easy-to-use and efficient tooling that supports the integration of semantic data
into modern (cloud) technologies.

\section{Conclusion and Further Research}

In this paper we present \textit{owl2proto}, an approach to bridge the world of
ontology design with the world of modern cloud native frameworks and services.
Our approach leverages protobuf, a popular format and framework for the
description of data as well as a binary representation of the actual transfer.
We show, using an example ontology based on cloud resources, that
\textit{owl2proto} addresses the requirements of seamlessly integrating semantic
data into the Cloud. 

We do note however, that there is further research to be conducted to fully
address requirements of adaptability, stemming from complex requirements of
protobuf itself. One possible approach could be to re-use techniques from the
efforts to standardize RDF dataset canonicalization. Furthermore, projects like
protovalidate\footnote{\url{https://github.com/bufbuild/protovalidate}} could be
used to translate OWL restrictions and constraints. Lastly, more validation of
the approach itself using real-world ontologies and use-cases needs to be
performed.

\section*{Acknowledgements}

This work was funded by the Horizon Europe project EMERALD, grant agreement ID
101120688.

\bibliographystyle{apalike}
{\small
\bibliography{example}}

\section*{\uppercase{Appendix}}
\label{appendix}


\begin{lstlisting}[language=protobuf3, style=protobufscript,label=lst:options,caption=The protobuf file defining our OWL options.]
message EntityEntry {
  string iri = 1;
  repeated string parent = 2;
}
message PropertyEntry {
  string iri = 1;
  repeated string parent = 2;
  string class_iri = 3;
}
message PrefixEntry {
  string prefix = 1;
  string iri = 2;
}
message Meta {
  repeated PrefixEntry prefixes = 1;
}
extend google.protobuf.MessageOptions {
  optional EntityEntry class = 50000;
}
extend google.protobuf.FieldOptions {
  optional PropertyEntry property = 50000;
}
extend google.protobuf.FileOptions {
  optional Meta meta = 50000;
}
\end{lstlisting}

\end{document}